# Some considerations about cosmogenic production of radioactive isotopes in Ar as target for the next neutrino experiments

Mihaela Pârvu*, I. Lazanu**

*University of Bucharest, Faculty of Physics, Bucharest-Magurele, POBox MG 11, Romania,*
* e-mail: mihaelaparvu3@gmail.com
** e-mail: ionel.lazanu@g.unibuc.ro

The concept of the LAr TPC technology that is an excellent tracking - calorimeter detector will be used for the next generations of neutrino experiments. In this class of detectors both the scintillation light emitted and the charge produced by the ionization are used to detect and identify the characteristics of the primary particle. The reduction of the radioactive background, the knowledge of the sources and mechanisms of its production as well as the characteristics of the signals have as consequence the increase of the sensitivity of huge detectors and the capability to discriminate between various particles interacting with the detector. Cosmogenic sources of background or activation of different materials become more important in this context. The radioactivity induced by cosmogenic reactions in Ar is discussed by considering muon capture and reactions induced by neutrons as sources of background. The simulated cross sections for the considered nuclear reactions are obtained using TALYS and EMPIRE codes, highlighting the similarities and differences between the results of these nuclear codes and the level of concordance with the few existing experimental data.

Key words: cosmogenic radioactivity, Ar, negative muon capture, neutron reactions.

## 1. INTRODUCTION

The good knowledge of the radioactive background and the search for materials that produce lower values for it is an important requirement for rare event experiments. Cosmogenic sources of background or activation of different materials in a detector system become more important with the increase of sensitivity of experiments and thus a good knowledge of unwanted processes is necessary. The main sources of background are: i) environmental radioactivity including geo-radioactivity, cosmic rays and their secondaries; ii) intrinsic radioactivity as radio impurities in the components of the detector and radio impurities in the shield materials; iii) activation of detector materials during exposure to radiation – in this class (α, n) reactions, neutrons from fission, as well as reactions induced by muons and neutrons.

Some comments about these sources of background must be done. The flux of high-energy muons induced by cosmic ray interactions decreases as depth increases while the angular dependence is due to the surface profile. Cosmic-ray muons themselves can be easily detected and vetoed, but muon induced spallation backgrounds, especially fast neutrons and long lifetime isotopes, are extremely dangerous for low background counting experiments. At low energies, neutrons are generated by α particles and fission processes of Uranium and Thorium in the rocks - this component depends on the site but is independent on depth. The muons could penetrate deep rocks and could produce capture reactions and this contribution is related to the site depth. Radioactive radon concentration in the air depends on local geology, but increases in closed halls. This can only be tuned by proper ventilation. Cosmogenic processes, the detector itself, concrete around the detector, supports, shielding, electrical connections, etc. may produce radioactive nuclides and thus contribute to the radioactive background.

In the last period both signals produced in the bulk of Ar detector, i.e. the scintillation light emitted in the vacuum ultraviolet (VUV) and the charge produced by the ionization are separately or simultaneously used to detect and identify the characteristics of the primary particle. Huge Ar detectors must have high sensitivity and the capability to discriminate between various particles interacting with the detector for efficient background reduction.

One of the next generations of neutrino experiments is based on the concept of the LAr Time Projection Chamber technology that is an excellent tracking - calorimeter detector. The pioneering work started with the successive prototypes for ICARUS (single phase technology): 0.003, 0.05, 3, 10 and 478 tons (T600) on surface and eventually to its underground commissioning. The alternative technology – the double phase Large Electron Multiplier (LEM) readout with charge extraction and amplification coupled to a very long drift path is in progress, small demonstrator is now operational, the prototype is in construction (with a volume ≈ 530 m$^3$ liquid argon and ≈ 70 m$^3$ gas) and final GLACIER design (20-100 kt) is technically feasible to be built [1].

Although external background from cosmic rays, i.e. hadronic components, is suppressed in the conditions of operating underground at high depth, muons with low and very low energies can survive the shielding of rocks and escape the trigger systems, or are generated by neutrinos and represent the ideal particles for capture processes in different materials, in particular in Ar. Also, the very high energy component of the flux of muons penetrating the shielding of rocks up to underground induces spallation reactions and thus activation of elements can be produced.

The neutrons existing in underground laboratories are those from astrophysical sources, those produced by reactions induced by atmospheric or solar neutrinos or those generated by cosmic muons that have penetrated so far. Neutrons and muons as well as atmospheric or solar neutrinos, geo-neutrinos, neutrinos from reactors are the major sources of external radioactive background, depending on the experimental configurations and materials used in walls, shielding, but also on the composition of the huge underground detector. Accidentally neutrinos from supernovae or other sources could be important.

In its turn, the scintillation signal is influenced by the radioactive background. As an incident particle will produce both singlets and triplet dimers, the scintillation is a product of the two radiative decays. The singlet decays quickly, being responsible for most of the prompt light seen in the scintillation spectrum, whereas the triplet decays with a longer lifetime. The time constant of the singlet and triplet decays have been measured in all phases. For the present case the singlet lifetime and the triplet lifetime, both in ns are: 7.0 ± 1.0 and 1600 ± 100 respectively [2]. The results of a spectroscopic study on liquid and gasous Ar [3] put in evidence the differences in the emission spectra as well as those due to the contributions of different impurities [4]. The LAr spectrum is dominated by an emission feature (126.8 nm) analogue to the 2nd excimer continuum in the gas phase, confirming the previous results of Doke [5]. Weak-emission features in the wavelength range from 145 to 300 nm can be observed. The structure at 155 nm in the gas phase has only a very weak analogue in the liquid phase. The structure at longer wavelengths up to 320 nm is addressed as the 3rd continuum emission in the gas phase. The scintillation yield of Ar for electron excitation is 40000 photons/MeV [6]. The presence of radioactive isotopes in Ar or other radioactive isotopes increases the number of photons emitted by scintillations. In addition, the impurities present in Ar (intrinsic or produced by cosmogenic processes) present particular emission features at specific wavelengths and must be investigated to improve particle discrimination capabilities of LAr detectors [3]. Usually wavelength shifters normally convert the VUV scintillation photons into ultraviolet (UV) or visible light where comparatively inexpensive photomultipliers can be used for light detection. All the spectral information which may be very important for particle identification is lost with this technique.

In this short contribution, details of the muon capture in $^{40}$Ar are discussed and the cross sections for possible production of radioactive isotopes by neutron activation are simulated using nuclear codes TALYS and EMPIRE.

**2. Negative muon capture in $^{40}$Ar**

Usually the energy loss of muons is parameterised in the form [7]:

$$\frac{dE_\mu}{dx} = -\alpha - \beta E_\mu$$

The critical energy is defined as $\varepsilon_\mu = \alpha/\beta$. Here the thickness of the crossed material is measured in g/cm$^2$. For standard rock, the typical values for the factors α and β are $\alpha \cong 2$ MeV g$^{-1}$cm$^2$ and $\beta \cong 4\times 10^{-6}$ g$^{-1}$cm$^2$. If the muon with initial energy $E_\mu^0$ penetrates a depth $h$, the average energy of a beam of muons is:

$$\langle E_\mu(h)\rangle = \left(E_\mu^0 + \varepsilon_\mu\right)^{-\beta h} - \varepsilon_\mu$$

and the minimum energy of a muon at the surface, necessary to penetrate the depth $h$ and obtaining the residual energy $E_\mu(h) \approx 0$ is: $E_{\mu,min}^0 = \varepsilon_\mu\left((\varepsilon_\mu)^{\beta h} - 1\right)$. In fact, this condition favours the nuclear capture process in nuclei. The capture process is:

$$(Z,A) + \mu^- \rightarrow (Z-1, N+1)^* + \nu_\mu$$

Analytical equations for the stopping rate of negative muons as well as production rate of radionuclides in capture processes can be found in Ref. [8]. For Ar the total capture rate is $(1.20 \pm 0.08) \times 10^6$ s$^{-1}$ [9]. In this particular case, the primary product is an excited state of either Cl or S nuclei, with energy from zero up to about 100 MeV [10]. If the excitation energy is below 6 MeV, this state de-excites electromagnetically, otherwise by nucleon emission. The results for nucleon emission are presented in Table 1.

*Table 1*
Muon capture in Ar

| Primary process | Final state | Isotopic yield/stopped muon [%] * | Details for the final radioactive nucleus ** | | Daughter is stable? | |
|---|---|---|---|---|---|---|
| | | | $T_{1/2}$ | Decay energy [MeV] | Yes | No |
| $_{18}^{40}Ar + \mu^-$ | $_{17}^{40}Cl + \nu_\mu$ | 7.29 ± 0.24 | 1.35 min | β$^-$: 7.480 | $^{40}Ar$ | |
| | $_{17}^{39}Cl + n + \nu_\mu$ | 49.05 ± 1.61 | 55.6 min | β$^-$: 3.442 | | $^{39}Ar$ |
| | $_{17}^{38}Cl + 2n + \nu_\mu$ | 15.45 ± 0.85 | 37.24 min | β$^-$: 4.917 | $^{38}Ar$ | |
| | $_{17}^{38}Cl^m + 2n + \nu_\mu$ | 1.61 ± 0.02 | 715 ms | IT: 0.671 | $^{38}Ar$ | |
| | $_{16}^{39}S + p + \nu_\mu$ | 0.21 ± 0.13 | 11.5 s | β$^-$: 6.640 | | $^{39}Cl$ |
| | $_{17}^{38}S + d / np + \nu_\mu$ | <1.2 | 170.3 min | β$^-$: 2.937 | | $^{38}Cl$ |

* Average values form more measurements; original data from [10].
** From [11]. These electrons produce a background in scintillations emission.

With a yield more than 67.31% per stopped muon, free neutrons are produced in the bulk of Ar and could initiate new reactions.

From the analysis of reaction products from different channels, it is clear that the $^{39}$Ar isotope appears with the highest direct yield from $^{39}$Cl disintegration and from subsequent disintegrations of $^{39}$S (49.23±1.62 yield/stopped muon). With a half life of $T_{1/2}$ =269 y and maximum beta decay energy of 0.565 MeV, the isotope $^{39}$Ar induce unwanted permanent background in scintillation and charge signals. Usually atmospheric Ar contains the radioactive $^{39}$Ar with a measured ratio $^{39}$Ar/$^{40}$Ar ≅ 1 x $10^{-16}$ g/g resulting in a specific activity of of 1 Bq/kg [12]. Any other source will increase this ratio. The isotope $^{39}_{18}Ar$ can be extracted from $^{40}$Ar by isotopic separation.

## 3. Cosmogenic reactions induced by neutrons in $^{40}$Ar

For the neutron spectrum, some parameterizations exist at sea level [13, 14]. For the underground case, it is necessary to obtain the transmitted flux, but also the component due to the process (μ,n) following the prescriptions of different authors or by simulating the interactions in the materials penetrated by muons. The production of fast neutrons from cosmic-ray muons was predicted by different authors, for example by Wang *et al*. [15] or Boehm *et al*. [16]. Malgin and co-workers investigated in a series of publications, see [17, 18, 19], the phenomenology of the muon induced neutron yield and thus neutron induced rate, obtaining an universal but empiric parameterization. In the general case, for a muon with the an energy $E_\mu$, the neutron yield in a material with mass number *A* is:

$$Y_\mu(A, E_\mu) = \frac{N_A}{A} \langle \sigma_{\mu A} \delta_n \rangle$$

where $N_A$ is Avogadro's number, $\langle \sigma_{\mu A} \delta_n \rangle$ is a mean value of the product of a μA-interaction cross-section and neutron multiplicity, $\delta_n$. In this equation, neutron yield is measured in (n/μ/(g/cm²)). For the muon energy range between about 40 GeV up to ~ 400 GeV, authors obtained an universal empirical formula for the muon induced neutron yield:

$$Y_n = \gamma E_\mu [\text{GeV}]^{\varepsilon_1} A^{\varepsilon_2}$$

Here the constants have the following values: $\gamma = 4.4 \times 10^{-7}$ cm²/g, $\varepsilon_1 = 0.78$, $\varepsilon_2 = 0.95$.

Another approach is represented by the Monte Carlo MUSIC and MUSUN simulation codes for muon transport through matter [20, 21] and this code is an usual instrument to estimate the rate to neutron production from muons in the underground.

Atmospheric neutrinos represent another source of background because they can interact with the detector producing muons and hadrons. Rein and Sehgal [22] describe the probable reactions with a single pion in the final state, muon and strange particles, in the resonance region up to *W* = 2 GeV

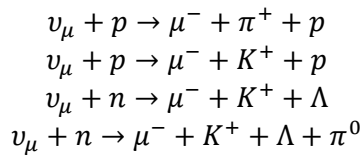

$$\nu_\mu + p \rightarrow \mu^- + \pi^+ + p$$
$$\nu_\mu + p \rightarrow \mu^- + K^+ + p$$
$$\nu_\mu + n \rightarrow \mu^- + K^+ + \Lambda$$
$$\nu_\mu + n \rightarrow \mu^- + K^+ + \Lambda + \pi^0$$

What is important to mention in this case is that these processes cannot be completly eliminated using trigger systems because the interactions could be produced in the bulk of the detector. Most of the background events in these channels are rejected because of the presence of other particles and kinematical configurations.

Radioactive isotopes produced in the activation reactions represent a different direction of investigation.

The isotopes of Ar: $^{35}$Ar, $^{37}$Ar, $^{39}$Ar, $^{41}$Ar, $^{42}$Ar and $^{43}$Ar are all radioactive. From all, $^{41}$Ar and $^{42}$Ar could be produced by neutron absorption, directly and as successive processes. Because the half-life of $^{41}$Ar is only 109.34 min, the production of $^{42}$Ar assumes high fluxes of neutrons.

The existence of a very small number of experimental data in Ar is a major problem in the analysis or estimation of the background. In this case we used the simulation of the nuclear reactions using TALYS and EMPIRE 3.2 codes to estimate the cross sections.

TALYS [23] is a software package for the simulation of nuclear reactions in the 1 keV - 200 MeV incident energy range for target mass numbers between 12 and 339 and considering neutrons, protons, deuterons, tritons, alphas, photons and hadrons as projectiles and produced particles. The code incorporates modern version of the nuclear models for elastic processes, direct reactions, pre-equilibrium reactions, compound reactions, fission reactions and a large nuclear structure database. As output information, the code calculates total and partial cross sections, energy spectrum, angular distributions, double-differential spectra, residual production cross sections and recoils. This code is frequently used by the scientific community for estimations of cross sections of interest in this field.

EMPIRE 3.2 is an alternative code for the simulation of nuclear reactions [24]. It is designed for calculations over a broad range of energies and incident particles. This code covers the same types of projectiles as TALYS and the energy range extends from the beginning of the unresolved resonance region for neutron-induced reactions ($\sim$ keV) and goes up to several hundred MeV for heavy-ion induced reactions. As in the case of TALYS, the major nuclear reaction mechanisms accounts. Details of the models implemented can be found in different papers, for example in [25]. This package contains the full EXFOR (CSISRS) library of experimental reaction data that are automatically retrieved during the calculations. In both cases only predefined parameters of the codes were used. These codes were chosen since they contain a wide spectrum of nuclear reaction models that provides the quantitative calculations. EMPIRE and TALYS are the only widespread all-in-one codes (meaning that all required reaction mechanisms are implemented in one software package) [26]. The results obtained using these codes allow a comparative analysis of the predicted cross sections, highlighting discrepancies between their results and the comparison with experimental data if available.

The elastic cross section (Figure 1) shows significant resonance behaviour at low energies below 650 keV. In the resonance energy region, the availability of measured data is essential for the correct understanding of the neutron - argon interaction.

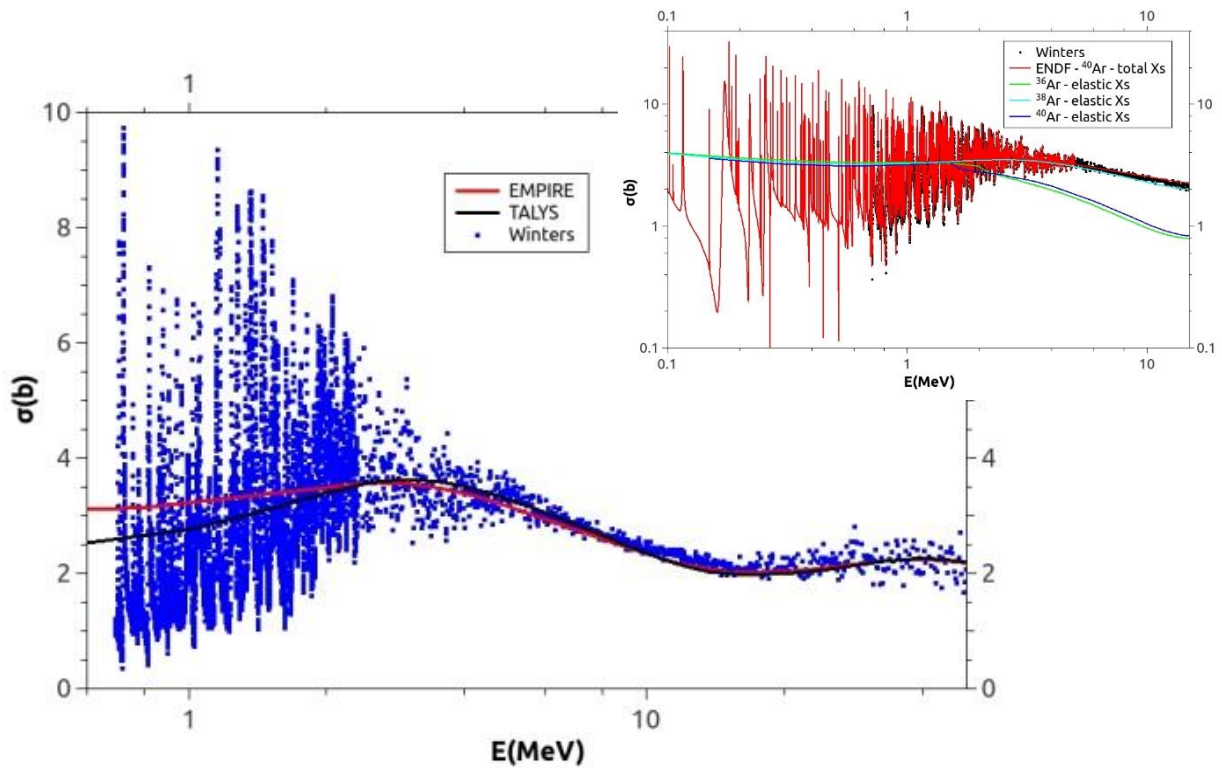

**Figure 1.** Results for cross sections as a function of the neutron energy for elastic $^{40}Ar(n, n)\,^{40}Ar$ using TALYS and EMPIRE codes. Elastic processes on $^{38}Ar$ and $^{36}Ar$ are also presented. Total cross section n - Ar is also included for reference.

For energies above the resonances, both simulation codes produce similar results and suggest lower cross sections for isotopes $^{38}$Ar and $^{36}$Ar. The differential cross section for elastic scattering of neutrons from Ar at 6.0 MeV was measured by Mac Mullin and co-workers [27]. Optical-model parameters for the elastic scattering reactions were determined from the best fit to these data and for total elastic scattering cross section and was found to differ by 8 % compared to a local optical-model for $^{40}$Ar, suggesting that new data are necessary for improving parameters used in the Monte Carlo models for simulations. Very good concordance between the two simulation codes was also obtained for n - Ar total cross sections. The results are presents in Figure 2.

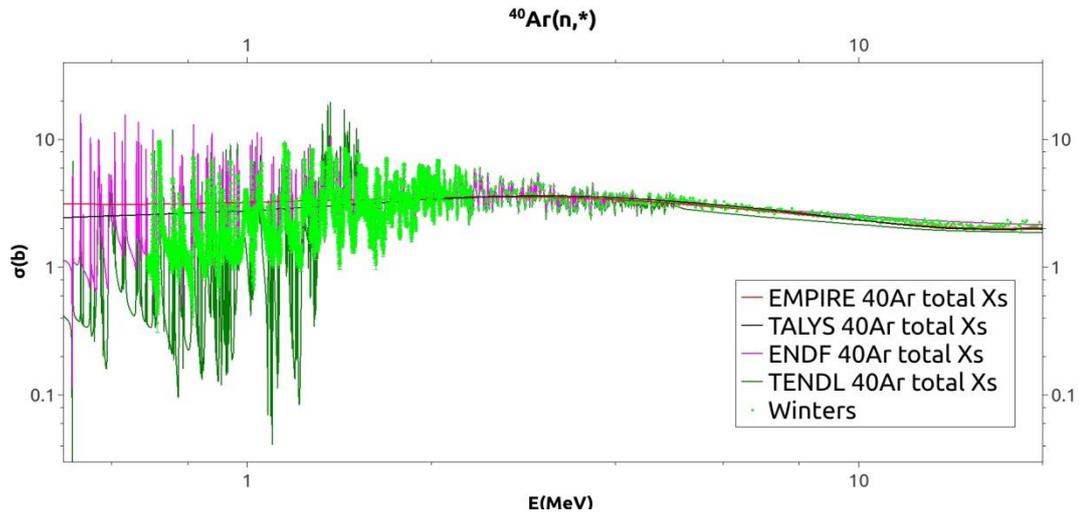

**Figure 2.** Results of for the total cross sections as a function of the neutron energy using TALYS and EMPIRE codes.

**Isotope** $^{39}_{18}Ar$. Its production is dominated by the process $^{40}Ar(n,2n)^{39}Ar$ which has an energy threshold of about 1 MeV [28]. Also, $^{39}Ar$ can be produced cosmogenically, by negative muon capture as discussed, or as capture on $^{39}K$, but this aspect is not of interest for the present discussion. We used both codes in order to calculate the specific energy dependence of the cross sections of interest in the production of the isotope $^{39}$Ar. The results are presented in Figure 3.

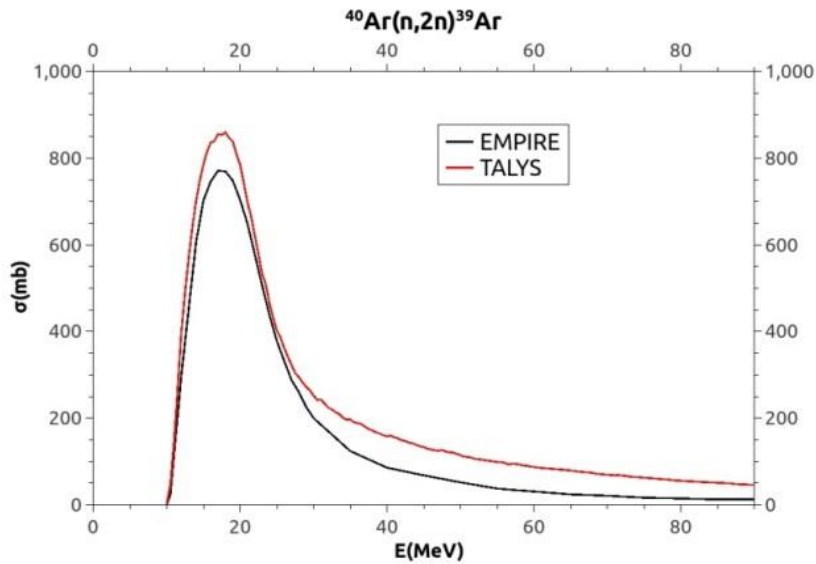

**Figure 3.** Energy dependence of the cross-section in the reaction (n, 2n) on $^{40}$Ar. TALYS and EMPIRE nuclear reaction codes are used.

The two codes give relatively similar results, with a maximum cross-section at the energy of 18 MeV when the calculations are made using TALYS and 17 MeV using EMPIRE. At the maximum, the EMPIRE results are about 100 mb values higher than those obtained using TALYS. For energies above 30 MeV the situation is reversed, and EMPIRE generates values higher with up to 50 mb. Unfortunately, there are no experimental data available.

**Isotope** $^{42}_{18}Ar$. The half-life of this isotope is 32.9 years and the beta decay of its daughter isotope, $^{42}$K, has the maximum electron energy of 3.52 MeV [29]. Recently, the concentration of $^{42}_{18}Ar$ in the Earth's atmosphere was estimated to be in the ratio of $(6.8^{+1.7}_{-3.2}) \times 10^{-21}$ atoms per atom of $^{40}$Ar [30]. A possible way to produce this isotope is a two steps neutron capture reaction in usual Ar: $^{40}Ar(n,\gamma)^{41}Ar$ and thus $^{41}Ar(n,\gamma)^{42}Ar$ or as a process induced by alpha particles: $^{40}Ar(\alpha,2p)^{42}Ar$.

The successive processes induced by neutrons were simulated using TALYS and EMPIRE. The results are presented in Figure 4. The experimental data for $^{40}Ar(n,\gamma)^{39}Ar$ reaction are from EXFOR library. The results for TALYS give relative confidence with experimental data while the EMPIRE code overestimates the data with a factor of around (5÷7). For the second reaction, experimental data do not exist.

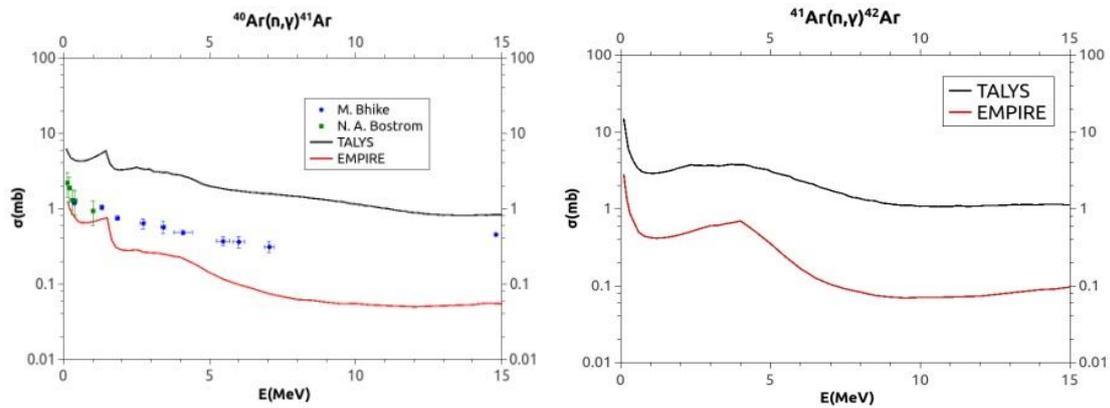

**Figure 4:** Results for the cross section as a function of the neutron energy for radiative - capture process (a) $^{40}Ar(n,\gamma)^{41}Ar$ and (b) $^{41}Ar(n,\gamma)^{42}Ar$ using TALYS and EMPIRE nuclear reaction codes. Experimental data from EXFOR, from [31; 32, 33] are also shown.

The alternative process induced by alpha particles was previously investigated [34, 35] and the maximum in the cross section is around 10 mb. This process was not considered in this contribution because it is improbable to produce alpha particles in the bulk of argon.

**Isotope $^{37}$S** can be produced as (n, α) reaction. The energy of the emitted β-particle equals 4.86 MeV. Its half-life is ~5.05 min. In this case, the cross-section calculated with EMPIRE fits better the available experimental data than the one obtained using TALYS – see Figure 5.

Discrepancies up to a factor of 2 between these two codes were obtained. For energies above 22 MeV the codes predict similar cross sections.

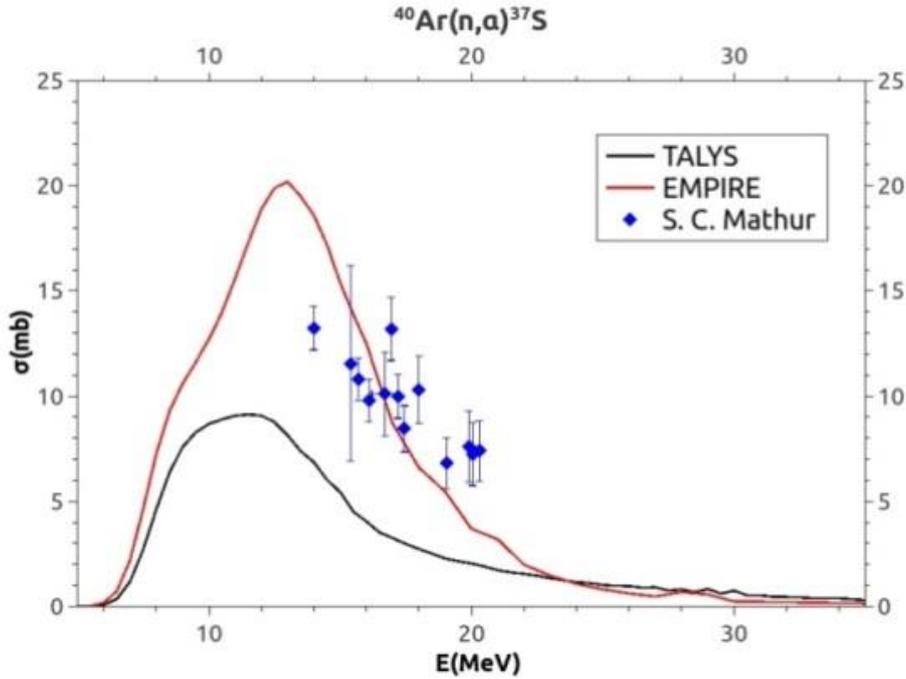

**Figure 5**: Cross section as a function of the neutron energy for $^{40}_{18}\text{Ar}(n,\alpha)^{37}_{16}\text{S}$ reaction using TALYS and EMPIRE codes; experimental data from EXFOR are also shown [36]

## 4. CONCLUSIONS

Ar is one of the most important materials used in neutrino experiments, being used as target and detector. One important direction of study is the investigation of the sources of production of radioactive isotopes inside Ar bulk. It is crucial to understand the associated radioactive contamination, especially the production of muons, neutrons and radioactive isotopes. In this contribution, negative muon capture in Ar as well as production of radioactive Ar are discussed. Even if the production rates calculated with these cross-sections are small, considering the large amount of liquid argon used as detector we can say that this kind of study is very important in avoiding the false signals that may appear in the detection system. These investigations are in progress.

## ACKNOWLEDGEMENTS

This work was partially supported by the Programme CERN-RO, under Contract 9/2017, NEPHYLAro.